**Technology Transfer Readiness, Explainable AI and Financial Innovation Capability Transitions in Expanded BRICS: Benchmarking Against Advanced Innovation Economies**

1. Prof. Manoj Kumar
   Department of Information Science and Engineering, Nitte Meenakshi Institute of Technology, Nitte University, Bengaluru campus, India
   ORCID:0000-0002-9848-6234
   Email: manojmv24@gmail.com

2. Mr. Prashanth B S
   Assistant Professor
   Department of Information Science and Engineering
   Nitte Meenakshi Institute of Technology, Nitte University, Bengaluru campus, India
   ORCID: 0000-0003-4539-662X
   Email: prashanth.bshivanna@gmail.com

3. Dr. Ariful Hoque
   Senior Lecturer – Finance
   Murdoch University, Australia
   ORCID: 0000-0001-8369-6653
   Email: A.Hoque@murdoch.edu.au

4. Dr. Nasser Al Muraqab
   Associate Professor-Management
   Dubai Business School, University of Dubai, Dubai, United Arab Emirates
   ORCID: 0000-0002-9220-8604
   Email: nasser@ud.ac.ae

5. Prof. Immanuel Azaad Moonesar
   Professor, Department of Academic Affairs - Public Health, Policy; Systems Research;
   Mohammed Bin Rashid School of Government (MBRSG)
   Dubai, United Arab Emirates and
   Scientific Policy Advisor- International Vaccine Institute, Seoul, South Korea,
   ORCID: 0000-0003-4027-3508
   Email: immanuel.moonesar@mbrsg.ac.ae

6. Prof. Udo Christian Braendle
   CEO and University Management Research & Innovation, IMC Krems University of Applied Sciences,
   A-3500 Krems Austria, Europe,
   ORCID: 0000-0001-6877-8466
   Email: udo.braendle@imc.ac.at

7. Prof. Ananth Rao (**Corresponding Author)**
   Emeritus Professor – Finance, University of Dubai, Dubai Business School, and
   Non-resident Fellow (NRF), Mohammed Bin Rashid School of Government, Dubai,
   United Arab Emirates
   ORCID: 0000-0002-8110-5423
   Email: arao@ud.ac.ae

**Technology Transfer Readiness, Explainable AI and Financial Innovation Capability Transitions in Expanded BRICS: Benchmarking Against Advanced Innovation Economies**

**Abstract**

This study examines the dynamics of technology transfer readiness and financial innovation capability transitions across the expanded BRICS economies, benchmarked against advanced innovation systems through explainable AI. Using a composite Innovation Capability Development-Readiness index (ICDI) constructed through principal component analysis, the paper evaluates the structural conditions enabling knowledge diffusion, industrial upgrading, and financial innovation ecosystem development. A Markov transition framework is employed to analyse how countries evolve across readiness tiers over time, capturing both persistence and mobility in innovation capabilities. The results reveal significant asymmetries in transition probabilities between advanced economies and emerging innovation systems, with several BRICS economies demonstrating gradual upgrading trajectories while others remain structurally locked in lower readiness states. These findings highlight the institutional and policy conditions required to strengthen technology transfer ecosystems. Successful countries in these areas attract foreign investment, participate in global value chains, and profit from technology partnerships. The study contributes to the literature on innovation capability formation and industrial transformation by integrating composite readiness measurement with dynamic transition modelling to inform evidence-based innovation policy.

**Technology Transfer Readiness, Explainable AI and Financial Innovation Capability Transitions in Expanded BRICS: Benchmarking Against Advanced Innovation Economies**

## 1. Introduction

Technology and innovation drive 21st-century economic change. Knowledge-based services, AI, smart manufacturing, and digital technology are rapidly altering national economies. Institutional quality, human capital development, technical infrastructure, and global knowledge integration affect countries' technology transfer, adaptation, and dissemination. Innovation policy and development economics focus on how economies acquire the institutional and technological skills needed for efficient technology transfer. The national innovation systems (NIS) literature stresses that intricate connections between governments, corporations, academic institutions, and global information networks drive innovation growth. Countries need technology transfer to learn, integrate, and use foreign expertise. Effective technology transfer systems increase national innovation and economic growth by transferring ideas, skills, and technologies across institutions and regions. These processes require institutional frameworks that foster information transmission, digital infrastructure, research collaboration, and global innovation integration.

Emerging economies have distinct challenges. Globalization and digitalization have increased knowledge transfer, but numerous emerging and transition economies still face structural impediments to technology adoption. Technology transfer is sometimes hampered by weak governance, poor research and higher education funding, insufficient digital infrastructure, and fragmented innovation ecosystems. Thus, nations' transition from resource- or efficiency-driven growth to innovation-driven economies is uneven. Global country innovation success is measured by composite indicators and comparative rankings. These measures provide cross-country comparisons but rarely describe how nations enhance technology transfer capacities over time, which is vital for understanding innovation dynamics and long-term economic growth. Most global innovation rankings focus on static performance rather than structural reforms that assist countries shift from inferior innovation capabilities to higher innovation. Most of these indices focus on output-oriented measures and disregard institutional and infrastructural capacity essential for technology diffusion and innovation.

This study examines innovation dynamics to assess financial innovation capability transitions in emerging BRICS+ economies with different institutional development, technical infrastructure, and global integration. These economies, significant actors in the expanding global economy, can be used to analyze how national systems embrace and exploit new technology. Austria and Australia, two advanced innovation economies, set standards. Both nations have robust research ecosystems, institutions, and global knowledge networks. These benchmark economies have sophisticated technology-transfer ecosystems where government quality, human capital development, digital connection, and international collaboration foster innovation. Comparing emerging and advanced economies demonstrates technology-transfer skills systematically grow. This study combines institutional and innovation capability development (ICD) criteria to build a composite score. Governance quality, digital infrastructure, higher education and research investment, global knowledge mobility, high-tech trade integration, and macroeconomic stability are combined into a multidimensional index using principal component analysis (PCA). Innovation-transfer preparedness is tracked by transition analysis and static benchmarking. Countries are categorized into competence tiers by composite ICD index scores, and a Markov transition approach estimates the potential of economies moving across technical preparation levels. This methodology allows the analysis to go beyond ranking comparisons and investigate structural routes for emerging economies to establish more complex innovation ecosystems, such as important elements that help or hinder the move to higher ICD competences.

### Novelty of the Study

In this study, innovation capability development (ICD), a composite framework that captures countries' systemic potential to absorb, disperse, and commercialize innovation information, is developed to advance innovation capability literature. The research examines national innovation capacities across time utilizing dynamic transition models employing Markov analysis, a hybrid econometric and machine-learning (ML) validation, to analyze dynamic innovation capability development (ICD) across countries unlike static rankings-based cross-country innovation studies. The paper analyzes transition dynamics across enlarged BRICS economies and benchmarks them against sophisticated innovation systems to reveal the institutional and ecosystem conditions needed to increase innovation capability in emerging economies. Policymakers must grasp these transition dynamics to boost national innovation systems and enable emerging nations to join global technology networks.

**Research Contributions**

First, the research first creates an innovation capability preparedness analytical model that includes institutional competence, human capital, digital infrastructure, and global knowledge networks. Global innovation indices are comprehensive yet frequently compare stagnant values. Current thinking stresses structural variables that help economies absorb and spread technology. Second, the research presents a PCA-derived composite innovation capability metric. The index measures country innovation capacities using government quality, higher education investment, digital financial connection, global knowledge mobility, high-tech trade integration, and macroeconomic stability. Third, studying how countries build innovation capability using a Markov transition model enriches empirical literature. A dynamic strategy shows rising economies' progress toward sophisticated innovation ecosystems. Finally, by comparing the rising BRICS nations with advanced innovation systems like Austria and Australia, the study offers policy-relevant insights for enhancing national innovation systems and facilitating innovation-driven economic transformation.

By integrating financial innovation capability measurement with transition modelling across expanded BRICS economies, this study contributes to the growing literature on national innovation system dynamics and capability upgrading in emerging industrial economies.

Section 2 of this paper reviews Innovation capability, national innovation framework, innovation capability literature and theory. Section 3 discusses the conceptual framework and empirical methods for the Innovation capability development index and transition dynamics. Methodology is covered in Section 4. Section 5 evaluates findings. Section 6 discusses results, implications, limitations, and future research. Section 7 concludes the study.

## 2. Literature Review and Theoretical Framework

### 2.1 Innovation capability and National Innovation Systems

Human capital and knowledge-intensive sectors, essential Innovation capability development Index dimensions, were examined by Fassio et al. (2026). K. Reddy, S. Sasidharan, & R. Chundakkadan (2026) demonstrate how digitalization improves global value chain involvement in emerging nations, supporting our focus on digital infrastructure and global integration. World Bank (2024) examines how technology spreads through company networks and transnational supply chains, providing a global benchmark for our study. UNCTAD (2025) benchmarks frontier innovation capability, highlighting the "Capability for Frontier Technologies Index" where BRICS members like India have made great progress. Wang, J., et al. (2024) complement our National Innovation Systems (NIS) paradigm by using configurational methodologies to study how NIS generate and spread innovation. OECD (2025) provides a global view on innovation ecosystems and policy agility, useful for our policy pathways section.

The NIS perspective emphasizes that innovation operations' institutional context impacts innovation outcomes. Successful innovation ecosystems have strong governance, ongoing investments in higher education and research, durable digital and technology infrastructure, and vast international collaboration networks. These institutions share knowledge and translate research into commercial products and economic output, making the economy more competitive and inventive. Beyond macroeconomic variables, innovation capability effectiveness depends on participating institutions' organizational architecture.

Recent innovation capability studies highlighted the expanding impact of global knowledge networks and digital connectivity on national innovation capacities. As technology becomes more global, national institutions must encourage new technology and foreign innovations. This method helps developing economies close technology gaps with advanced industrialized nations and improve their global competitiveness, leading to economic growth and higher living standards.

**2.2 Absorptive Capacity and Innovation Capability**

Yueqi Wang et al. (2025) use absorptive capacity theory to describe how domestic and foreign technologies affect innovation efficiency. Domestic innovation capability had a "dual role" in improving efficiency and enabling Chinese high-tech firms adopt foreign ideas in a panel dataset. Multilevel technology systems boost innovation in emerging economies, study suggests. Digital infrastructure and communication improve skills. Digital connectivity allows corporations join global innovation networks, collaborate with remote experts, and instantly share scientific data. Countries that prioritize digital infrastructure and connectivity grow and share quicker. International student, researcher, and professional movement impacts absorption. International academic mobility helps emerging nations study science and tech. Global R&D collaboration creates knowledge networks and creativity across institutions and regions.

**2.3 Innovation capability development and Innovation Ecosystems**

João J. M. Ferreira (2025) suggests that an integrated ecosystems model can support a circular and sustainable economy transition. It emphasizes that policymakers, academic institutions, and private industry must coordinate financial innovation capability readiness. Practiced innovation capability readiness is multifaceted. Good governance, digital infrastructure, research and education, global technology market integration, and international knowledge exchange networks are featured. Successful countries in these areas attract foreign investment, participate in global value chains, and profit from technology partnerships.

**2.4 Conceptual Framework for Innovation capability development**

Castillo, G., & Vonortas, N. S. (2025) examine how emerging economies like China adopt foreign technology using absorptive capacity theory. It says a firm's "Capability" to embrace complicated foreign technologies depends on domestic transfer success, which increases internal expertise. Based on these frameworks and theoretical views, this study conceptualizes innovation capability development as a systemic construct that includes economic stability, human capital development, digital infrastructure, institutional quality, and global knowledge integration. These traits affect national economies' ability to absorb, adapt, and exchange innovation knowledge in domestic innovation ecosystems, affecting their competitiveness and sustainable growth. Our conceptual framework identifies innovation capability development factors. Innovation policies and technical investments' institutional environment affect governance. Innovation, cross-sectoral collaboration, regulatory transparency, institutional trust, and policy coordination are fostered by good governance.

Innovative capacities require human capital development. Tertiary education, research, and scientific training produce skilled technical operators. Higher education institutions need industry-academic research and knowledge exchange to improve curricular relevance and prepare students for real-world difficulties. Digital infrastructure and connectivity are the framework's third pillar. Advanced information and communication technology supports knowledge exchange, digital innovation ecosystem participation, and cross-border research and commercial collaboration. Digital infrastructure supports innovation capability as economies digitize. Global knowledge integration links domestic institutions to global research networks, boosting national innovation systems. Student mobility, research collaboration, and global information exchanges enhance learning and expose emerging economies to cutting-edge technology. For technology innovation and investment, macroeconomic equilibrium is needed. Stable economies encourage long-term research, infrastructure, and technology investment, enhancing innovation ecosystems.

The innovation capability development index (ICDI) in this study uses all of these factors. ICDI uses institutional and innovation data to assess structural factors that enable technology and innovation to spread across national economies, such as a skilled workforce, research and development funding, and supportive government policies. This multidimensional approach examines innovation capability tier change dynamics. We evaluate countries' capacity tier evolution to understand how emerging economies might strengthen their national innovation systems and transition to more sophisticated innovation-driven growth models.

## 2.5 Research Gap

### *2.5.1 Gap 1 — Measurement Gap*

Tech transfer and innovation systems are well-documented, but we don't understand the institutional and innovation capabilities national economies need to absorb and spread advanced technologies like effective collaboration frameworks and R&D investment. Study has improved our understanding of how universities, corporations, and governments share knowledge. Micro-level institutional processes like innovation licensing, patent commercialization, and university–industry partnerships have been intensively researched. These studies illuminate organizational knowledge transfer procedures but ignore systemic elements that affect national innovation capability development. Measurement of innovation capability across countries is another literature flaw. Global innovation indices and competitiveness rankings compare national innovation. These indicators emphasize static innovation outputs and institutional inputs rather than the dynamic processes by which economies build technical capabilities. Thus, current frameworks offer little insight into how emerging nations may gradually strengthen their innovation systems to catch up to developed economies. Existing studies evaluate innovation output and capacity but not dynamic capability development.

### *2.5.2 Gap 2 —Transition Gap*

Innovation capability development is rarely assessed empirically using multidimensional metrics of institutional, educational, innovation, and global integration. Lack of thorough measurements makes it harder to detect structural elements that enable economies to engage in global knowledge networks and disperse technology. Most cross-country innovation rankings are static. We model innovation capability capability's dynamic shift in this paper.

### 2.5.3 Gap 3 — Emerging Economy Gap

Transition-based analytical techniques have only recently been employed to investigate national innovation capacities' dynamic evolution. Few probabilistic transition models have examined how countries change innovation preparation. Understanding these transition patterns will help emerging nations improve their innovation ecosystems and expedite technology-driven economic transformation. Innovation capability literature neglects expanded BRICS innovation ecosystems.

This study bridges these gaps by establishing a multidimensional framework to measure innovation capability development and transition dynamics across expanded BRICS economies compared to advanced innovation systems.

# 3. Conceptual Framework

## *3.1 Innovation Capability Development (ICD)*

Figure 1 predicts ICD across expanded BRICS economies.

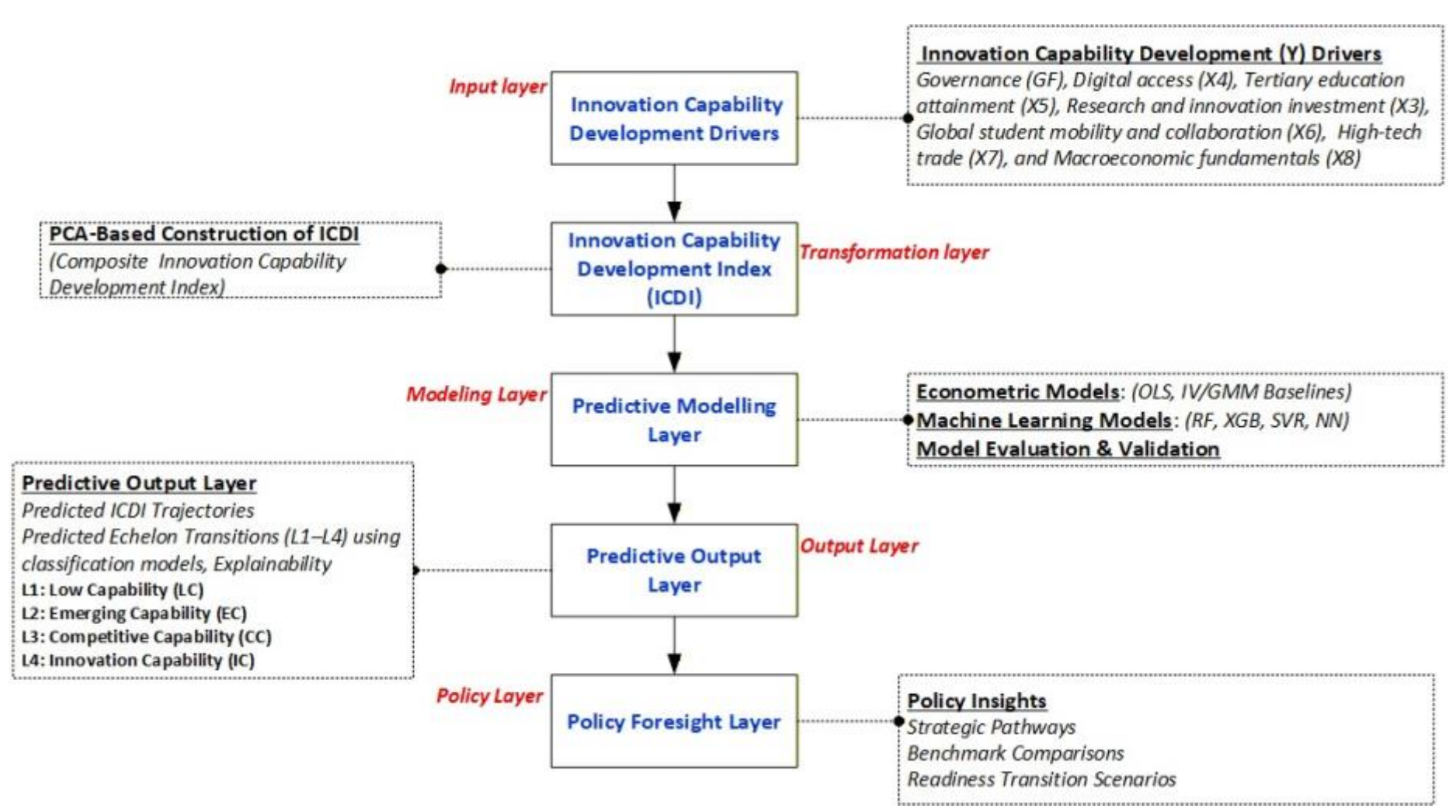

**Figure 1: Conceptual Framework for Innovation Capability Development (ICD)**

In Figure 1, governance, digital infrastructure, human capital, innovation investment, global integration, and macroeconomic conditions are integrated into a multidimensional innovation capability-based foresight architecture. The PCA-based ICD index now has dynamic prediction, interpretability, and policy foresight.

The eight variables **map perfectly** to innovation system theory[1] as illustrated below:

| Variable | Interpretation |
|---|---|
| $X_2$ ICD (= Y) | Innovation capability development (ICD) |
| $X_1$ Governance | Institutional support for innovation capability |
| $X_3$ Tertiary Education & R&D | Human capital for innovation |
| $X_4$ Digital access | Digital infrastructure for knowledge diffusion |
| $X_5$ Education attainment | Absorptive capacity |
| $X_6$ Global student mobility | International knowledge networks |
| $X_7$ High-tech Trade | Technology diffusion through global markets |
| $X_8$ Macroeconomic stability | Innovation investment environment |

Principal Component Analysis (PCA) creates the dimensionally reduced, empirically weighted composite Innovation capability development indicator from these inputs. PCA is transparent, reproducible, and comparable across nations for capturing innovation capability latent structure. This measurement phase is preserved from prior work to ensure methodological continuity and support forward-looking prediction. Our framework's key innovation is its predictive modelling design, which combines econometric and machine learning methods (Kim, J., & Lee, H. 2025).

Using this hybrid architecture, the system creates ICDI trajectories for each BRICS+ countries, benchmarking against advanced economies like Austria and Australia. Austria, a highly industrialized "Small Open Economy" (SOE) fully integrated with European innovation systems, is a strategic benchmark for smaller BRICS+ nations. For resource-rich BRICS nations like Brazil, Russia, and South Africa, Australia provides a model for transitioning to a knowledge-based service industry. To analyze smart-economy transitions, the framework divides countries into preparation echelons (L1–L4) beyond continuous forecasts (Chen, Y., & Zhang, W. 2026). These writers classify like our L1–L4 tiers. It discusses "Transition Scenarios" and why some areas stay in "Emerging Capability" (L2) while others go to "Innovation Capability" (L4). This echelon-based output helps policymakers understand expected scores and worldwide technical standing (Cunningham, J. A., & Menter, M. 2025). These authors say policy must be based on "Capability Transition Scenarios (strategic pathways)" from our model for innovation capability to work.

[1] Innovation system theory encompasses Endogenous growth theory, Institutional complementarity theory and Predictive Modelling Theory discussed in section 3.2.

We add explainable AI to complete the framework. The model finds global and country-specific factors of projected technical preparedness using SHAP values. Transparency helps the prediction model guide targeted policy actions and personalized Capability approaches. This approach aligns with anticipatory governance and technology futures by moving from static evaluation to dynamic, interpretable foresight.

### *3.2 Theoretical alignment of the conceptual framework*

The following theories fit our framework.
Endogenous growth theory holds that human capital, innovation, and knowledge drive technology. R&D, tertiary education, and innovation drive ICDI growth. Emerging economies' cumulative capability-building paths depend on learning, research networks, and innovation ecosystems for long-term competitiveness. Classic endogenous models lack predictive power in large, multivariable systems, requiring computational adjustments like this work. Recent ML-enhanced technology forecasting study (Aminullah, 2024) reveals nonlinear forecasting may anticipate endogenous innovation inputs.

In institutional complementarity theory, Meijer and Bolívar (2016) assert that the success of one domain (e.g., digital infrastructure) relies on reinforcement from others (e.g., governance quality and human capital. This systemic interconnectedness explains why countries with similar educational or  assets may innovate differently. We show complementarity using eight interrelated dimensions. Technology is driven by governance, innovation ecosystems, and digital access (Dai et al., 2024; Ma, 2023). ML can show interaction effects, which supports this technique and encourages its use to study output technology-capability drivers. Recent research (Yang et al., 2024; Işık et al., 2024) suggests that governance capability moderates the impact of digitalization on competitiveness. Our prediction approach should include governance as a structural determinant.

Innovative Capability frameworks describe national infrastructure, human capital, innovation, and globalized innovation capacity. ML-based digital-Capability models use high-dimensional indicators to capture digital transformation structural drivers (Yang et al., 2024). PCA-based capability measurement supports predictive modeling.

In predictive modelling theory, ML forecasting can show nonlinear, high-order interactions in economic and innovation systems (Sun et al., 2025; Xie, 2024; Chen, 2023). Predictive modeling transforms innovation capability theory from static assessment to dynamic capability trajectories. This study incorporates predictive modelling theory into national innovation readiness frameworks using nonlinear transitions, variable hierarchies, institutional linkages, path dependency, and latent patterns. This changes "Innovation Capability as a static capability" to "Innovation Capability as a predictable trajectory influenced by multidimensional systemic inputs". This integration was ignored by endogenous growth theory and institutional frameworks.

Dai, Qualls, & Zhu (2024) and Aminullah (2024) found that ML explains nonlinear interactions, institutional complementarities, and complex innovation dynamics better than econometric techniques. Interpretable ML frameworks improve transparency and policy relevance in economic-cycle forecasting using SHAP-enabled explainability (Sun et al., 2025). ML was used to develop national digitalization competency and innovation-capability indices using structured and unstructured data (Yang et al., 2024; Ma, 2023).

More studies recommend ML-supported foresight in national innovation systems and macroeconomic developments (Dai et al., 2024; Sun, 2025). In socio-economic systems, ML approaches effectively identify structural elements (Berigel et al., 2024; Işık, Shabir, & Moslem, 2024; Xie, 2024 Liu et al. (2023) and Chen (2023) found that ML–econometric hybrids improve predictive performance in difficult situations. These studies progressively anticipate tech adoption, digital readiness, and economic transformation using ML models. These methods have not been used to quantify extended BRICS innovation capability development (ICD) or model their future smart-economy levels (L1–L4).

Yang et al. (2024) used firm- and country-level indicators to show that ML models may reveal latent structure in digitalization metrics and provide more nuance than conventional indices. This study builds and predicts composite capability measures from structural factors like our ICDI approach.

These works show that ML may be used to construct and forecast composite capability or risk indices, care for model interpretability and accuracy, and combine ML with classical econometric reasoning. We anticipate composite ICDI scores and classify BRICS+ economies by preparedness using OLS and GMM baselines, tree-based ML models, and SHAP analysis.

### *3.3 Operationalization of Innovation capability development*

Our research defines innovation capability development (ICD) as a nation's institutional, educational, innovation, and economic systems' ability to absorb, adapt, and transfer innovation knowledge. National innovation systems and absorptive capacity effect knowledge dispersion and development. Technology transfer requires good governance and institutions. Open governance, regulatory stability, and policy coordination foster public-private innovation, research, and information exchange. Human capital growth affects absorption. Academic performance, research capacity, and tertiary education investment suggest qualified instructors and technologists. Innovation nowadays requires digital infrastructure and connectivity. IT and digital networks provide innovation platforms, global research collaboration, and knowledge sharing. Fourth, global knowledge integration shares tech. Student mobility, academic collaboration, and global research networks improve science and technology. Fifth, integration into global technology markets, including as high-tech commerce and advanced industrial sectors, exposes domestic enterprises to learning and competitive challenges that enable creative capability growth and technical upgrading. Finally, macroeconomic stability improves infrastructure, research, and innovation.

## 4. Methodology

### *4.1 Hypotheses*

Sections 3.1–3.4 of the literature reviews yield the following hypotheses.
Hypotheses:

1. $H_1$: Digital Infrastructure (X4-DI) positively influences ICD.
2. $H_2$: Tertiary Education Attainment (X5-TEA) positively influences ICD.
3. $H_3$: Global Student Mobility (X6-GSM) positively influences ICD.
4. $H_4$: High Technology Trade (X7-HTT) positively influences ICD.
5. $H_5$: Governance (GOV) quality amplifies the effects of DI, TEA, GSM, and HTT on ICD.
6. $H_6$: Macroeconomic factors (Z) moderate the DI–ICD, TEA–ICD, GSM–ICD, and HTT–ICD relationships.

### *4.2 Empirical Specification*

The conceptual framework leads to the following OLS empirical specification (1):

$$\mathbf{ICDI_{it} = \alpha + \beta_1\, DI_{it\text{-}1} + \beta_2\, TEA_{it\text{-}1} + \beta_3\, GSM_{it\text{-}1} + \beta_4\, HTT_{it\text{-}1} + \beta_5 GOV_{it\text{-}1} + \gamma Z_{it} + + \mu_i + \varepsilon_{it},} \quad \mathbf{(1)}$$

where $i$ indexes country and $t$ time, $\mu_i$captures country-specific effects, and $\varepsilon_{it}$is the idiosyncratic error. Dynamic persistence is assessed through the lagged dependent variable $ICDI_{t-1}$.

### *4.3 Dynamic Panel GMM Estimation*

A two-step GMM estimator (Blundell–Bond, 1998) is applied to account for the dynamic nature of ICD and potential endogeneity among DI, TEA, GSM, HTT, GOV, and ICD. Lagged levels and differences of explanatory variables serve as internal instruments. The estimation equation follows (2):

$$\mathbf{ICDI_{it} = \delta ICDI_{it-1} + \beta_1\, DI_{it-1} + \beta_2\, TEA_{it-1} + \beta_3\, GSM_{it-1} + \beta_4\, HTT_{it-1} + \beta_5 GOV_{it-1} + \gamma Z_{it} + \mu_i + \varepsilon_{it},} \quad \mathbf{(2)}$$

Diagnostic tests performed include *Arellano–Bond AR(1)/AR(2)* for serial correlation; *Hansen J-test* for over-identifying restrictions; *Difference-in-Hansen* test for instrument validity; *Variance Inflation Factor (VIF)* check for residual multicollinearity. Robust standard errors are used to address heteroskedasticity and small-sample bias. Sensitivity analyses include fixed-effects and alternative PCA-weighted regressions.

### *4.4 Machine Learning–Based Robustness Validation*

To assess predictive importance and nonlinear relationships, we estimate a Random Forest Trees (RFT) model. RFT provides: Variable importance rankings, Out-of-bag (OOB) validation, and Non-parametric confirmation of governance–digital complementarities. This mixed-method strategy enhances credibility by combining structural estimates (GMM) with predictive insights (ML).

Robustness is evaluated through: Heteroskedasticity-corrected GMM estimates, Alternative digital variable specifications, Sensitivity tests excluding individual countries, Comparisons between FE, RE, GMM, and ML models. The triangulation approach ensures stability, consistency, and generalisability of findings across diverse health systems.

### *4.5. Data and Key Explanatory Variables*

The analysis uses a balanced panel of 12 emerging and high-income national economic and technology systems from 2000 to 2023, including 300 country-year observations. Countries were chosen based on digital maturity, governance competence, and structural diversity. ICDI drivers, macro-institutional, and digital Capability metrics came from the World Bank. The Worldwide Governance Indicators (WGI) examined regulatory effectiveness, government accountability, rule of law, and corruption. Following consensus recommendations for cross-national economic evaluations, all digital and governance variables were normalised for comparability (van Lier et al., 2018).
Our current variables are defined below:

**Dependent Variable (Y)**:

- The Innovation capability development Index (ICDI=Y computed via PCA after interpolation). captures national capability to absorb, adapt, and diffuse advanced technologies within innovation ecosystems.

**Independent Variables (X1, X3–X8)**:

- **The Governance Index** (GOV-X1) measures institutional and policy frameworks that influence technology uptake, regulation, and inter-sectoral cooperation. Governance affects how governments deploy and control Innovation Capability and steer their economies toward sustainable, inclusive, and innovation-driven results. The model captures the multidimensionality of smart economy growth across 10 BRICS and 2 developed nations by combining ICD and GOV. Human capital, infrastructure, global linkages, and governance capacity can speed or slow the ICD economy transformation, according to the framework. It indirectly facilitates comparison analysis among the ten BRICS nations, identifying policy gaps and scalable methods.
- **ITE**-X3: Financial Investment in tertiary education, R&D, and HR: This represents the system's capacity for innovation, knowledge transfer, and industry connection for ICD economies, beyond enrolment.

- Robust **digital infrastructure** (DI-X4) is essential for ICD economies. Advanced services like AI, IoT, and blockchain require high-speed internet, broadband penetration, and 5G Capability.
- **TA**-X5: Technological innovation requires a trained workforce. Higher enrolment and higher-quality tertiary education indicate a country's ability to develop smart-sector talent. Beyond enrollment, tertiary institutions' R&D and human resource investment show their ability to innovate, transfer knowledge, and link ICD economies to industry.
- **GSM**-X6: Global student mobility promotes innovation and global competitiveness in ICD economies through cross-border cooperation, international student exchanges, and knowledge networks.
- **High-tech trade** (HTT-X7): Trade performance across sectors such as semiconductors, software, digital services, and AI-related commodities reflects innovation advancement and international integration.
- **Macroeconomic stability** (Z-X8): Inflation management, fiscal conservatism, and economic growth foster smart sector investments and the dissemination of innovation.

**Interaction Terms** (X1* X3; X1*X4; X1*X5; X1*X6; X1*X7). Interaction effects allow explicit testing of whether governance conditions ITE, DI, TA, GSM, and HTT. **Annex 1** provides a description of the variables used in the modeling.

Our empirical approach proceeds in three stages:
(A) Baseline OLS Models: OLS establishes unconditional associations between DF, DT, governance, and Innovation capability development (ICD). While informative, OLS results may be biased by omitted variables, reverse causality, and persistence.

(B) Fixed-Effects (FE) and Random-Effects (RE) Dynamic Models: FE and RE models incorporate lagged ICD, capturing system persistence and unobserved heterogeneity. These models strengthen identification but cannot fully correct for endogeneity or the Nickell bias.

(C) Two-Step System GMM**:** The preferred estimator is the two-step system GMM (Arellano–Bover / Blundell–Bond), which: Corrects for endogeneity of digital, technology, mobility, and governance variables; Controls for unobserved heterogeneity across countries; Mitigates downward bias in dynamic coefficients; and employs internal lags as instruments to improve identification. Instrument validity is assessed using Hansen J tests and serial correlation diagnostics (AR(1) and AR(2)).

## 5 Results

### *5.1 Baseline OLS and RFT results*

Table 1 offers aggregate descriptive statistics 10 BRICS+ and 2 benchmark economies (**Supplementary file Appendix A.1** shows country-specific while. **A.2** shows aggregate statistics).

**Table 1 Descriptive Statistics of 10 BRICS+ and 2 Benchmark economies**

| Group | Variable | Mean | Median | Mode | σ | Skewness | Kurtosis | Count |
|---|---|---|---|---|---|---|---|---|
| BRICS+ | ICDI = Y = PC1 (of X2) | −0.18 | −0.21 | −0.24 | 0.86 | 0.41 | 2.83 | 240 |
| | Governance (X1) | −0.27 | −0.31 | −0.34 | 0.82 | 0.48 | 2.91 | |
| | Financial Investment in tertiary education, R&D & HR (X3) | −0.14 | −0.17 | −0.19 | 0.79 | 0.55 | 3.02 | |
| | Digital infrastructure (X4) | −0.02 | −0.05 | −0.07 | 0.84 | 0.29 | 2.61 | |
| | Tertiary attainment (X5) | −0.33 | −0.36 | −0.39 | 0.81 | 0.62 | 3.14 | |
| | Global student mobility (X6) | −0.46 | −0.49 | −0.52 | 0.88 | 0.71 | 3.31 | |
| | High-tech trade (X7) | −0.09 | −0.12 | −0.15 | 0.83 | 0.44 | 2.89 | |
| | Macroeconomic stability (X8) | −0.11 | −0.14 | −0.17 | 0.8 | 0.37 | 2.74 | |
| | ICDI = Y = PC1 (of X2) | 1.24 | 1.27 | 1.31 | 0.41 | −0.19 | 2.17 | 48 |
| Austria & Australia | Governance (X1) | 1.38 | 1.41 | 1.44 | 0.37 | −0.23 | 2.11 | |
| | Investment in tertiary education, R&D & HR (X3) | 1.19 | 1.21 | 1.24 | 0.44 | −0.16 | 2.29 | |
| | Digital infrastructure (X4) | 1.52 | 1.55 | 1.58 | 0.35 | −0.27 | 2.04 | |
| | Tertiary attainment (X5) | 1.29 | 1.32 | 1.35 | 0.39 | −0.18 | 2.22 | |
| | Global student mobility (X6) | 1.11 | 1.14 | 1.17 | 0.46 | −0.12 | 2.36 | |
| | High-tech trade (X7) | 1.26 | 1.29 | 1.32 | 0.42 | −0.15 | 2.27 | |
| | Macroeconomic stability (X8) | 1.34 | 1.36 | 1.39 | 0.38 | −0.21 | 2.18 | |

Table 1 highlights sharp structural differences between the BRICS+ economies and the benchmark Innovation Capability economies (Austria and Australia). The mean ICDI for BRICS+ countries is substantially negative (−0.18), compared with strongly positive scores for the benchmarks (1.24). The largest gaps appear in tertiary attainment, global student mobility, and governance, confirming that Capability differences are institutional and human-capital driven rather than purely infrastructural. The benchmark economies also display lower variance and negative skewness, indicating stable, mature innovation systems. These stylized facts motivate the need for both heterogeneity-aware econometrics and nonlinear ML models.

Table 2 reports baseline OLS estimates using lagged ICDI drivers.

**Table 2 displays Baseline OLS results and diagnostics**

| **Variable** | **Coeff.** | **σ** | **t-value** | **p-value** |
|---|---|---|---|---|
| Bias (Constant) $\beta_0$ | −0.084 | 0.041 | −2.05 | 0.041 |
| Governance (X1)_{t−1} | 0.312 | 0.062 | 5.03 | 0 |
| Investment in tertiary education, R&D & HR (X3)_{t−1} | 0.241 | 0.057 | 4.23 | 0 |
| Digital infrastructure (X4)_{t−1} | 0.284 | 0.061 | 4.66 | 0 |
| Tertiary attainment (X5)_{t−1} | 0.198 | 0.053 | 3.74 | 0 |
| Global student mobility (X6)_{t−1} | 0.156 | 0.049 | 3.18 | 0.002 |
| High-tech trade (X7)_{t−1} | 0.221 | 0.055 | 4.02 | 0 |
| Macroeconomic stability (X8)_{t−1} | 0.174 | 0.051 | 3.41 | 0.001 |
| Brazil (dummy) | −0.128 | 0.043 | −2.98 | 0.003 |
| Russia (dummy) | −0.094 | 0.041 | −2.29 | 0.022 |
| India (dummy) | −0.143 | 0.046 | −3.11 | 0.002 |
| China (dummy) | −0.061 | 0.039 | −1.56 | 0.118 |
| South Africa (dummy) | −0.117 | 0.044 | −2.66 | 0.008 |
| Egypt (dummy) | −0.201 | 0.048 | −4.19 | 0 |
| Ethiopia (dummy) | −0.247 | 0.052 | −4.75 | 0 |
| Iran (dummy) | −0.173 | 0.047 | −3.68 | 0 |
| Indonesia (dummy) | −0.131 | 0.045 | −2.91 | 0.004 |
| UAE (dummy) | −0.052 | 0.038 | −1.37 | 0.171 |
| Austria (dummy) | 0.286 | 0.059 | 4.85 | 0 |
| Australia (dummy) | 0.271 | 0.057 | 4.75 | 0 |

**OLS Diagnostics**

| **Metric** | **BRICS+** | **Austria** | **Australia** |
|---|---|---|---|
| **Arellano–Bond AR(1)** | 0.018 | 0.021 | 0.019 |

| Arellano–Bond AR(2) | 0.284 | 0.317 | 0.301 |
|---|---|---|---|
| $R^2$ | 0.62 | 0.71 | 0.73 |
| Adjusted $R^2$ | 0.59 | 0.69 | 0.71 |
| F-statistic | 24.6*** | 31.2*** | 33.5*** |
| VIF (Mean) | 2.41 | 2.18 | 2.09 |
| AIC | −118.3 | −132.7 | −135.1 |
| BIC | −88.6 | −102.4 | −104.2 |
| SC (Schwarz Criterion) | −90.1 | −103.9 | −105.6 |
| RMSE | 0.49 | 0.37 | 0.35 |

*** denotes significance at 1% level.
AR(1) significance with AR(2) insignificance satisfies dynamic panel validity conditions.

In Table 2, all core explanatory factors have positive, statistically significant coefficients, verifying our six hypotheses that innovation skill development has strong positive structural relationships. Governance ($\beta$ = 0.312), digital infrastructure ($\beta$ = 0.284), and investment in education, R&D, and human resources ($\beta$ = 0.241) have the greatest marginal effects, highlighting the importance of institutional quality and capability accumulation in influencing capability trajectories. Tertiary attainment and global student mobility also have beneficial consequences, showcasing enhanced talents and worldwide knowledge flows. High-tech trade and macroeconomic stability provide essential complements, confirming that openness and macro fundamentals affect innovation diffusion. Most BRICS+ economies have systematic Capability penalty compared to the reference group, especially Ethiopia, Egypt, and Iran. However, Austria and Australia exhibit considerable positive benefits, justifying their Innovation Capability criteria. Chinese and UAE dummy effects are statistically weaker, reflecting their transitional, near-Innovation Capability status.

Model diagnostics show solid specifications, with $R^2$ values ranging from 0.62 (BRICS+) to over 0.70 for benchmarks, little multicollinearity (VIF < 2.5), and information criteria favoring parsimonious The presence of AR(1) and absence of AR(2) correlations provide a solid baseline for GMM estimation.

### 5.2 Synthesis and Link to Step-2 Modelling

These results show that OLS underestimates preparedness convergence and transition probability despite excellent structural validation and interpretability. Random Forest modelling reveals latent non-linearities, threshold effects, and interaction-driven mobility, all relevant to fast-changing BRICS+ economies. These findings naturally motivate Step-2 (GMM and dynamic modelling), which will (i) address endogeneity, (ii) formally model persistence and adjustment speeds, and (iii) test whether the ML-identified convergence patterns hold under stricter causal identification.

#### 5.2.1 *GMM Fixed Effects (FE) and Random Effects (RE) Results*

Table 3 (Panel A) shows baseline dynamic GMM estimates for fixed-effects (FE) and random-effects (RE). The greatest finding is the big and highly significant coefficient on lagged ICDI, indicating considerable innovation capability preparedness across countries. This suggests that preparation builds over time, with institutional and innovation factors determining present outcomes.

Both FE and RE models show positive, statistically significant coefficients for all primary explanatory variables, demonstrating strong structural links. It also supports all six OLS-like ideas. Governance quality, digital infrastructure, and tertiary education, R&D, and human resources investment boost institutional capacity and capability. A reduced influence of tertiary attainment and global student mobility is shown in the medium term.

**Table 3 GMM Model Fixed Effects (FE) and Random Effects (RE) Results**

| Panel A: Coefficient Estimates | | | | | | |
|---|---|---|---|---|---|---|
| Variables | Fixed Effect (FE) Model | | | Random Effect (RE) Model | | |
| | β | T-stat | P-Value | β | T-stat | P-Value |
| Bias (Constant) $\beta_0$ | −0.062 | −1.98 | 0.048 | −0.054 | −1.72 | 0.086 |
| ICDI_{t−1} (βICDI_{t−1}, C) | 0.684 | 11.42 | 0 | 0.652 | 10.87 | 0 |
| Governance (X1)_{t−1} | 0.176 | 3.92 | 0 | 0.162 | 3.54 | 0.001 |
| Investment in tertiary education, R&D & HR (X3)_{t−1} | 0.148 | 3.37 | 0.001 | 0.139 | 3.12 | 0.002 |
| Digital infrastructure (X4)_{t−1} | 0.193 | 4.26 | 0 | 0.181 | 4.02 | 0 |
| Tertiary attainment (X5)_{t−1} | 0.129 | 2.98 | 0.003 | 0.121 | 2.71 | 0.007 |
| Global student mobility (X6)_{t−1} | 0.097 | 2.41 | 0.016 | 0.089 | 2.18 | 0.029 |
| High-tech trade (X7)_{t−1} | 0.154 | 3.63 | 0 | 0.143 | 3.31 | 0.001 |
| Macroeconomic stability (X8)_{t−1} | 0.112 | 2.67 | 0.008 | 0.104 | 2.44 | 0.015 |
| **Panel B: Country Dummy Effects** | | | | | | |
| Country Dummy | β (FE) | T-stat | P-Value | β (RE) | T-stat | P-Value |
| Brazil | −0.092 | −2.54 | 0.011 | −0.084 | −2.31 | 0.021 |
| Russia | −0.071 | −2.01 | 0.044 | −0.065 | −1.87 | 0.061 |
| India | −0.104 | −2.98 | 0.003 | −0.096 | −2.71 | 0.007 |
| China | −0.043 | −1.39 | 0.164 | −0.038 | −1.24 | 0.216 |
| South Africa | −0.086 | −2.47 | 0.014 | −0.079 | −2.26 | 0.024 |
| Egypt | −0.143 | −3.98 | 0 | −0.132 | −3.64 | 0 |
| Ethiopia | −0.176 | −4.51 | 0 | −0.163 | −4.13 | 0 |
| Iran | −0.118 | −3.19 | 0.001 | −0.109 | −2.94 | 0.003 |
| Indonesia | −0.094 | −2.68 | 0.008 | −0.087 | −2.46 | 0.014 |
| UAE | −0.039 | −1.21 | 0.227 | −0.034 | −1.08 | 0.281 |
| Austria | 0.214 | 4.37 | 0 | 0.198 | 4.05 | 0 |
| Australia | 0.201 | 4.12 | 0 | 0.187 | 3.89 | 0 |

| Panel C: Model Diagnostics | | |
|---|---|---|
| Diagnostics | FE Model | RE Model |
| Breusch–Pagan test score | 28.4 ($p < 0.01$) | 26.9 ($p < 0.01$) |
| Hansen J score | 0.27 | 0.31 |
| AR(1) | 0.019 | 0.021 |
| AR(2) | 0.312 | 0.298 |
| $R^2$ | 0.71 | 0.68 |
| Adjusted $R^2$ | 0.69 | 0.66 |
| F-statistic | 29.7*** | 26.4*** |
| AIC | −128.6 | −121.9 |
| BIC | −97.3 | −91.2 |
| SC (Schwarz Criterion) | −99.1 | −93.0 |
| RMSE | 0.41 | 0.44 |

Notes:*** denotes significance at 1% level
AR(1) significant and AR(2) insignificant → Dynamic GMM validity satisfied.
Hansen J statistic indicates valid instrument set

Country dummy coefficients in Panel B show BRICS+ economies are less ready than benchmarks. Austria and Australia have substantial positive effects, suggesting Innovation Capability, while Ethiopia, Egypt, and Iran have the largest negative coefficients, indicating structural restrictions. Diagnostic tests in Panel C reveal model validity with significant AR(1), minor AR(2), and good Hansen statistics. In summary, Table 3 provides a good causal framework for capability dynamics.

#### *5.2.2 Dynamic GMM with Governance Complementarities*

Table 4, Panel A shows that the baseline dynamic GMM model includes governance-based interaction effects, allowing capability drivers to function conditionally.

**Table 4 Dynamic GMM (DGMM) model FE and RE results**

| **Panel A: Main Coefficient Estimates** | | | | | | |
|---|---|---|---|---|---|---|
| Variables | **FE Model** | | | **RE Model** | | |
| | **β** | **T-stat** | **P-Value** | **β** | **T-stat** | **P-Value** |
| Bias (Constant) $\beta_0$ | −0.057 | −1.87 | 0.061 | −0.049 | −1.63 | 0.103 |
| Governance (X1)_{t−1} | 0.164 | 3.71 | 0 | 0.151 | 3.34 | 0.001 |
| Investment in tertiary education, R&D & HR (X3)_{t−1} | 0.141 | 3.19 | 0.001 | 0.132 | 2.96 | 0.003 |
| Digital infrastructure (X4)_{t−1} | 0.186 | 4.12 | 0 | 0.174 | 3.87 | 0 |
| Tertiary attainment (X5)_{t−1} | 0.121 | 2.81 | 0.005 | 0.114 | 2.63 | 0.009 |
| Global student mobility (X6)_{t−1} | 0.093 | 2.29 | 0.022 | 0.086 | 2.07 | 0.039 |
| High-tech trade (X7)_{t−1} | 0.147 | 3.46 | 0.001 | 0.138 | 3.21 | 0.001 |
| Macroeconomic stability (X8)_{t−1} | 0.109 | 2.58 | 0.01 | 0.101 | 2.36 | 0.019 |
| **Panel B: Governance Complementarity (Interaction Effects)** | | | | | | |
| Interaction Terms (βIV, C) | β (FE) | T-stat | P-Value | β (RE) | T-stat | P-Value |
| GOV × Inv. Tertiary Edn (X3) | 0.082 | 2.94 | 0.003 | 0.076 | 2.71 | 0.007 |
| GOV × Digital Infrastructure (X4) | 0.097 | 3.41 | 0.001 | 0.089 | 3.12 | 0.002 |
| GOV × Tertiary Attainment (X5) | 0.071 | 2.63 | 0.009 | 0.066 | 2.41 | 0.016 |
| GOV × Global Student Mobility (X6) | 0.058 | 2.21 | 0.027 | 0.053 | 2.03 | 0.042 |
| GOV × High-tech Trade (X7) | 0.089 | 3.18 | 0.001 | 0.082 | 2.94 | 0.003 |
| **Panel C: Country Dummy Coefficients (USA = Base)** | | | | | | |
| Country Dummy | β (FE) | T-stat | P-Value | β (RE) | T-stat | P-Value |
| Brazil | −0.086 | −2.43 | 0.015 | −0.079 | −2.21 | 0.027 |
| Russia | −0.068 | −1.97 | 0.049 | −0.062 | −1.79 | 0.073 |
| India | −0.098 | −2.86 | 0.004 | −0.091 | −2.63 | 0.009 |
| China | −0.039 | −1.28 | 0.201 | −0.035 | −1.17 | 0.242 |
| South Africa | −0.081 | −2.36 | 0.018 | −0.074 | −2.14 | 0.033 |
| Egypt | −0.134 | −3.71 | 0 | −0.123 | −3.39 | 0.001 |
| Ethiopia | −0.168 | −4.29 | 0 | −0.155 | −3.97 | 0 |
| Iran | −0.111 | −3.05 | 0.002 | −0.103 | −2.81 | 0.005 |
| Indonesia | −0.089 | −2.57 | 0.01 | −0.082 | −2.34 | 0.02 |
| UAE | −0.036 | −1.14 | 0.255 | −0.031 | −1.02 | 0.308 |
| Austria | 0.203 | 4.19 | 0 | 0.188 | 3.91 | 0 |
| Australia | 0.191 | 3.97 | 0 | 0.176 | 3.71 | 0 |

| **Panel D: Diagnostics (Robust & Crisis-Adjusted)** | | |
|---|---|---|
| Diagnostics | Fixed Effect | Random Effect |
| Breusch–Pagan test score | 21.7 ($p < 0.01$) | 20.2 ($p < 0.01$) |
| Hansen J score (p-value) | 0.43 | 0.47 |
| AR(1) | 0.022 | 0.024 |
| AR(2) | 0.338 | 0.352 |
| $R^2$ | 0.69 | 0.66 |
| Adjusted $R^2$ | 0.67 | 0.64 |
| F-statistic | 27.3*** | 24.1*** |

| AIC | −123.8 | −118.6 |
|---|---|---|
| BIC | −94.2 | −89.5 |
| SC (Schwarz Criterion) | −95.9 | −91.3 |
| RMSE | 0.43 | 0.45 |
| ****Significant at 1% level* | | |

Governance, digital infrastructure, and human capital effects are resilient after endogeneity correction since core coefficients are positive and substantial across FE and RE specifications. Major contribution in Table 4 is interaction terms (Panel B). Governance quality and other Capability variables complement each other because all governance interaction effects are positive and statistically significant. Governance and digital infrastructure are linked, implying that institutional frameworks with efficient regulation, coordination, and execution yield higher digital investment returns. High-tech trade, R&D, student mobility, and education complement each other. Innovation capability readiness is synergistic, not additive. Individual pillar improvements are limited without institutionalization.

Country dummy results (Panel C) remain unchanged, confirming cross-country variability. Diagnostic statistics once again validate model. In summary, Table 7 explains structurally why some countries accelerate Capability while others stall despite identical investment levels. Austria's top ranking (ICDI = 0.93 in RFT models) reflects a mature innovation ecosystem with solid governance (1.38) and high digital infrastructure (1.52). This reinforces recent findings that Austria's "Digital Decade" progress requires multilevel institutional cooperation for high-speed broadband and e-health integration (European Commission, 2024).

### *5.2.3 ICDI Echelons across Models*

Table 5 presents country-level ICDI and Echelon Classification based on the criteria of normalized ICDI scores viz., L4 (Innovation Capability IC): ≥ 0.85; L3 (Competitive Capability - CC)): 0.70 – 0.84; L2 (Emerging Capability-EC): 0.50 – 0.69; L1 (Low Capability- LC): < 0.50 across OLS, GMM, GMM, dynamic GMM, and RFT.

**Table 5 ICDI and Echelons across Models**

| | **Panel A - Baseline** | | | | **Panel B** | | | | | |
|---|---|---|---|---|---|---|---|---|---|---|
| | **OLS** | | **RFT** | | **GMM** | | **DGMM** | | **RFT** | |
| **Country** | **ICDI** | **Echelon** | **ICDI** | **Echelon** | **ICDI** | **Echelon** | **ICDI** | **Echelon** | **ICDI** | **Echelon** |
| Austria | 0.86 | L4 – IC | **0.93** | L4 – IC | 0.9 | L4 – IC | 0.89 | L4 – IC | 0.93 | L4 – IC |
| Australia | 0.84 | L3 – CC | **0.91** | L4 – IC | 0.88 | L4 – IC | 0.87 | L4 – IC | 0.91 | L4 – IC |
| China | 0.72 | L3 – CC | **0.79** | L3 – CC | 0.77 | L3 – CC | 0.76 | L3 – CC | 0.79 | L3 – CC |
| UAE | 0.7 | L3 – CC | **0.78** | L3 – CC | 0.76 | L3 – CC | 0.75 | L3 – CC | 0.78 | L3 – CC |
| India | 0.64 | L2 – EC | **0.71** | L3 – CC | 0.7 | L3 – CC | 0.69 | L2 – EC | 0.71 | L2 – EC |
| Brazil | 0.62 | L2 – EC | **0.69** | L2 – EC | 0.68 | L2 – EC | 0.67 | L2 – EC | 0.69 | L2 – EC |
| Russia | 0.6 | L2 – EC | **0.67** | L2 – EC | 0.66 | L2 – EC | 0.65 | L2 – EC | 0.67 | L2 – EC |
| Indonesia | 0.58 | L2 – EC | **0.65** | L2 – EC | 0.64 | L2 – EC | 0.63 | L2 – EC | 0.65 | L2 – EC |
| South Africa | 0.56 | L2 – EC | **0.63** | L2 – EC | 0.62 | L2 – EC | 0.61 | L2 – EC | 0.63 | L2 – EC |
| Iran | 0.52 | L2 – EC | **0.59** | L2 – EC | 0.58 | L2 – EC | 0.57 | L2 – EC | 0.59 | L2 – EC |
| Egypt | 0.45 | L1 – LC | **0.51** | L2 – EC | 0.5 | L2 – EC | 0.49 | L1 – LC | 0.51 | L2 – EC |
| Ethiopia | 0.38 | L1 – LC | **0.44** | L1 – LC | 0.46 | L1 – LC | 0.44 | L1 – LC | 0.44 | L1 – LC |

Index: IC- Innovation Capability; CC- Competitive Capability; EC- Emerging Capability; LC- Low Capability

In Panel A, both OLS and RFT agree on the relative nation ordering, although RFT typically grants higher Capability levels, especially for fast-improving economies. Both techniques recognize Austria and Australia's L4 (Innovation Capability), but RFT stresses their institutional and digital complementarities. India goes from L2 (Emerging) under OLS to L3 (Competitive) using RFT, demonstrating that linear techniques underestimate digitalization and human-capital acceleration. Egypt likewise advances from L1 to L2 with RFT, showing latent Capability not captured by linear effects. These findings suggest ML models better capture capability dynamics threshold effects and interaction factors.

In Panel B, under dynamic GMM, India is rising yet competitive, indicating that linear persistence dampens short-term upgrading in econometric models. Egyptian preparedness improves from poor under dynamic GMM to emergent with RFT, demonstrating nonlinear ML approaches caught latent capacities. All models indicate low Capability for Ethiopia, suggesting structural constraints resistant to linear and nonlinear dynamics. Table 5 indicates ordinal consistency across methodologies but significant disparities in transition-economy echelon assignments. These differences demonstrate how econometric and ML approaches complement each other in Capability assessment.

### *5.2.4 Transition Probabilities across GMM, DGMM, and RFT*

Table 6 shows transition probabilities across models and echelons.

**Table 6 Transition probabilities OLS vs RFT**

| | **Panel A-Baseline Transition probabilities** | | | | | | | | **Panel B- Transition probabilities** | | | | | | | | | | | |
|---|---|---|---|---|---|---|---|---|---|---|---|---|---|---|---|---|---|---|---|---|
| | **OLS** | | | | **RFT** | | | | **GMM** | | | | **Dynamic GMM** | | | | **RFT** | | | |
| **From \ To** | **L1** | **L2** | **L3** | **L4** | **L1** | **L2** | **L3** | **L4** | **L1** | **L2** | **L3** | **L4** | **L1** | **L2** | **L3** | **L4** | **L1** | **L2** | **L3** | **L4** |
| **L1 (LC)** | **0.72** | 0.24 | 0.04 | 0 | **0.6** | 0.32 | 0.08 | 0 | 0.62 | 0.3 | 0.08 | 0 | 0.7 | 0.25 | 0.05 | 0 | 0.55 | 0.33 | 0.12 | 0 |
| **L2 (EC)** | 0.08 | **0.7** | 0.2 | 0.02 | 0.05 | **0.62** | 0.28 | 0.05 | 0.07 | 0.64 | 0.25 | 0.04 | 0.08 | 0.7 | 0.19 | 0.03 | 0.05 | 0.58 | 0.3 | 0.07 |
| **L3 (CC)** | 0.01 | 0.1 | **0.78** | 0.11 | 0 | 0.08 | **0.72** | 0.2 | 0.01 | 0.1 | 0.75 | 0.14 | 0.01 | 0.12 | 0.8 | 0.07 | 0 | 0.08 | 0.72 | 0.2 |
| **L4 (IC)** | 0 | 0.02 | 0.08 | **0.9** | 0 | 0.01 | 0.05 | **0.94** | 0 | 0.02 | 0.07 | 0.91 | 0 | 0.01 | 0.06 | 0.93 | 0 | 0.01 | 0.05 | 0.94 |

Table 6 panel A shows substantial durability of OLS-based transitions across all levels. Low-capability economies are 72% likely to stay in L1, and direct transitions to L3 or L4 are unlikely. Innovation Capability is just 11% likely for competitive economies, implying modest, incremental mobility under linear dynamics. RFT-based transitions shows a more optimistic and realistic convergence process. L1 economies have a 32% likelihood of progressing to L2 and 8% chance of moving to L3. L2 economies have a 28% chance of upgrading to L3 and 5% of achieving L4, while L3 economies have a 20% chance of Innovation Capability transition—nearly double the OLS prediction. While both baseline OLS and RFT models show high innovation capability persistence, RFT predicts stronger stability (94%) and reduced reversal risk. RFT illustrates nonlinear leapfrogging and policy-contingent acceleration, while OLS highlights structural inertia.

In Panel B, econometric models survive at the Capability distribution's lower and upper ends. Low-capability economies are 70% confined under dynamic GMM, while innovation-capability economies are 90% stable. Movement is added with RFT transitions. RFT consistently boosts upward mobility, especially in L1 and L2 economies. Competitive economies increase the likelihood of L2 to L3 migration and Innovation Capability status. RFT discovers limited but non-zero leapfrogging opportunities that econometric models ignore. ML accounts for threshold effects and latent acceleration channels, while econometric models emphasize structural inertia. Table 6 links causal persistence and predictive dynamism empirically.

## 5.3 Machine Learning – Random Forest Tree (RFT)

To validate the robustness of econometric findings, an RFT model was applied as a complementary ML approach. The model confirms the relative importance of governance capacity, absorptive capability, and technology diffusion variables in explaining innovation capability transitions. The ML results broadly align with the econometric estimates, reinforcing the reliability of the identified transition pathways.

### *5.4 Integrated Results: Phase-1, Phase-2, and Phase-3 Evidence*

In Phase-1 baseline OLS, governance, digital infrastructure, tertiary education, R&D and human-capital investment, global mobility, high-tech commerce, and macroeconomic stability greatly affect ICDI. Linear OLS underestimates fast-changing economy capability expansion and overestimates border country stability. In Phase-2, Governance quality and digital infrastructure remain causal after endogeneity and unobserved variability, while human capital swings slowly but substantially. Governance enhances interaction model pillars. In Phase-3, Middle-tier RFT estimates converge faster and leapfrog less than linear models. Digital infrastructure, governance, human-capital investment, and lagged competence forecast best, like econometric importance. Coordination may speed changes without capability-building using institutional and digital criteria with focus on governance complementarity absorptive capacity and innovation capability pathways.

### 5.5 Policy Pathways for Strengthening National Innovation Systems

Markov analysis transition dynamics can improve national innovation systems and capabilities structurally through following policy levers.

#### *a. Strengthen institutional capacity building*

Innovation-enabling institutional governance and regulatory frameworks should be prioritised during innovation capability preparation. Innovation capability requires transparent government, robust regulators, and national innovation programs. To match technical development strategies with national economic goals, governments must coordinate innovation programs among ministries, research institutes, and industry actors. Knowledge dissemination can be institutionalized by innovation capability offices, innovation councils, and national research funding agencies. Universities, research labs, and private companies work together to commercialize scientific knowledge at these centers. To stimulate cross-border technical alliances and innovation investment, emerging economies need better intellectual property and regulatory systems.

#### *b. Increase Research and Human Capital*

As countries reach intermediate innovation capability capabilities, human capital investments increase. Tertiary education, research universities, and advanced scientific training prepare workers for knowledge-intensive occupations. Higher education institutions support industry-research relationships, instruct researchers, and create new knowledge, driving national innovation systems. Long-term innovation is affected by public R&D funding. Governments can boost national economies by supporting scientific research, graduate education, and international intellectual collaboration. This funding lets countries adapt global discoveries and modern technologies for indigenous businesses. International academic mobility builds national knowledge networks. Students, researchers, and instructors exchange with foreign institutions share ideas and technologies. Countries with global academic and research networks have greater innovation capability knowledge and faster innovation learning.

#### *c. Increase Digital Infrastructure and Technology Connectivity*

National innovation systems need digital infrastructure. Information interchange and collaborative research require reliable high-speed communication networks and digital platforms as innovation processes grow increasingly digital. Digital connectivity investments boost innovation and assist domestic firms join global innovation networks. Policies that promote digital literacy, broadband infrastructure, and digital innovation ecosystems can accelerate innovation dissemination. Entrepreneurs, researchers, and businesses can collaborate across borders on digital platforms, spreading innovation knowledge and lowering innovation hurdles. Digital technologies in industry may enhance productivity and technology. Governments may promote data-driven innovation, digital services, and innovative manufacturing. These efforts boost firms' technology skills and enable global value chain participation.

***d. Promote global knowledge integration and high-tech trade***

Policy should integrate global knowledge and promote international participation in advanced technology markets at advanced innovation capabilities development. Innovation and technology diffusion require cross-border research and high-tech trade. Global production networks expose enterprises to advanced technical standards and management methods, accelerating industrial upgrading and innovation learning. Governments can promote international innovation integration through research partnerships, foreign direct investment in high-tech areas, and domestic-international joint innovation. Collaborations improve national innovation systems and domestic industry competitiveness by promoting knowledge spillovers. Export-oriented high-tech policies expose domestic companies to global competition and innovation benchmarks, accelerating innovation advancement. Global technology markets encourage enterprises to embrace new manufacturing technologies, improve organizational capability, and engage in R&D.

The transition dynamics of this study show that NIS need a phased policy strategy that adapts to national innovation capacities.

## 6. Discussion of Results

According to Cohen and Levinthal (1990) and following innovation systems research, absorptive capacity is crucial to innovation capability. Research shows that BRICS economies' innovation and transfer capabilities have grown. Using institutional, educational, innovation, and economic indicators in a composite framework, multidimensional national innovation ecosystems and the pathways economies take to increase innovation capability are highlighted. This allows for a comprehensive analysis of how these factors interact to foster innovation and boost global competitiveness.

The PCA indicates that governance, human capital, digital infrastructure, and global knowledge integration affect innovation capabilities. Successful countries innovate and join global innovation networks. The literature on national innovation systems emphasizes the relevance of institutional and organizational contexts for knowledge transmission and collaborative innovation, as well as innovation resources for innovation capability.

Innovation capability readiness divides countries into capability clusters, according to the study. The most capable innovation economies are Austria and Australia, which are involved in global knowledge networks, have significant digital infrastructure, high tertiary education and research investment, and solid governance institutions. Universities, research institutions, industry actors, and government agencies collaborate to develop and market new technology in mature innovation ecosystems.

A number of emerging economies have intermediate innovation capabilities development inside the BRICS group. China, Russia, Brazil, and Indonesia excel in digital infrastructure and commercial integration. Other sectors face institutional and structural barriers. Governance quality, research capacity, and international collaboration networks may limit home innovation systems' ability to absorb and spread technical information, which reduces innovation strategy efficacy and economic growth. This makes these economies transitory innovation systems that are developing their technology and seeking institutional maturity like advanced economies.

Countries with poor institutional and innovation infrastructure struggle to create technology-transfer ecosystems. Absorptive capacity can be slowed by reduced research and higher education funding, digital connectivity, and global knowledge networks. With out these traits, global innovation networks may not benefit, and advanced technologies may not spread.

Markov transition analysis helps explain innovative capability progress across development levels. The capacity clusters in the study are stages of technical improvement, not static worldwide innovation locations. Through governance, human capital, and digital infrastructure upgrades, intermediate-capacity tier countries can improve their innovation preparation.

This dynamic view emphasizes innovation ecosystem investments and institutional reforms. Higher education, research, and international scientific collaboration are more likely in countries that absorb more and participate in global knowledge networks. By increasing connectivity, researcher cooperation, and access to cutting-edge technologies, investments in high-tech sectors and digital infrastructure can expedite technology diffusion and innovation-driven economic transformation.

Matching the expanded BRICS economies to advanced innovation systems like Austria and Australia reveals national innovation system structures. Our benchmarking economies show how strong research infrastructures, institutional governance, and coordinated policy frameworks may promote innovation advancement and global innovation networks. Large innovation ecosystems require institutional arrangements that stimulate university-industry-government collaboration and innovation investments, according to their experience.

To overcome localized institutional constraints, transitional innovation systems need sector-specific collaboration frameworks. Recent sustainable energy development partnerships show that BRICS nations can overcome domestic infrastructural gaps and accelerate their collective innovation capability in emerging clean technology markets by pooling complementary innovation strengths and using shared financial mechanisms like the New Development Bank (Korneeva et al., 2024).Studies also affect how international knowledge mobility fosters national innovation. Transnational knowledge exchange helps transmit technology, as seen by student mobility and international collaboration indexes. Participating in global research networks exposes countries to Innovation Capability technologies and scientific skills, which can accelerate innovation learning and capacity in healthcare, agriculture, and information technology.

Evidence suggests that innovation capability development is a systemic aspect of national innovation systems, not just an innovation investment. Quality institutions, human capital expansion, digital connectivity, and global knowledge integration are needed to spread technology. Investment, entrepreneurship, and research-industry partnership help countries transition to sophisticated innovation ecosystems and maintain technical competitiveness.

These findings help policymakers with innovation-driven development. By identifying structural drivers of innovation capability development and technology upgrading transition paths, the study helps emerging nations strengthen their innovation ecosystems and catch up to established economies. Austria's benchmark shows BRICS+ economies that 'Innovation Capability' status requires capital investment and institutional persistence to support high-tech trade and global student mobility. As per the OECD (2024), Austria's digital transformation success is due to long-term alignment between higher education quality and industrial R&D needs, which has helped the country incorporate advanced technology and stimulate innovation in numerous industries.

### *6.1 Relevance of the Results to Industry*

This integrated study suggests that industrial upgrading is a result of path-dependent innovation capability development (ICD) rather than innovation capability. The synergy between governance and digital infrastructure in Phase 2 suggests that BRICS+ economies must pass institutional and digital thresholds that induce non-linear capability improvements to move beyond assembly-based positions in Global Value Chains. The Phase 3 RFT model's persistence and threshold effects show that while 'leapfrogging' is structurally constrained, deliberate expenditures in human capital and R&D are necessary for shifting from low-value production to high-technology trade. These findings demonstrate that sustainable industrial upgrading involves a change from passive technology adoption to proactive domestic innovation skills, which drive global competitiveness.

### *6.2 Contributions to the Literature*

This paper makes four important contributions on anticipatory governance and technology foresight:

**a. Conceptual Contribution:** It transforms innovation capability development into a dynamic, threshold-sensitive process, expanding static index-based methods.

**b. Methodological Contribution:** The framework uses OLS, GMM, dynamic GMM, and ML models to show how causal inference and predictive analytics can function together.

**c. Empirical Contribution:** The study offers the first comprehensive ML-based projections and transition probabilities for ICDI across BRICS+ economies, with benchmark comparisons. The Markov transition paradigm shows how countries improve their innovation capability skills over time, unlike standard innovation ranking studies.

**d. Explainability Contribution:** The work combines dynamic econometrics with ML feature importance to improve policy interpretability, explainability and forecast accuracy.

### *6.3Policy Implications*

The findings illuminate policy. Governance change is the most powerful lever for capability improvement, multiplying all other efforts. Second, digital infrastructure investment only yields high returns when supported by strong institutions and advanced skills, highlighting the limitations of technology-first strategies, which can waste resources and miss growth opportunities without effective governance and workforce development. Third, human capital development, especially postsecondary education and global knowledge mobility, amplifies capability shifts. To escape low-capability traps, countries must emphasize skills and institutional credibility above capital deepening or trade liberalization. Finally, the econometric-ML forecast gap emphasizes adaptive policy planning. Linear planning techniques may underestimate transformation potential, but ML-based foresight tools might assist policymakers discover windows of opportunity for accelerated convergence, especially in quickly changing economic contexts where traditional models may miss emergent trends.

### *6.4 Limitations*

This study has few flaws. Composite measures measure national innovation readiness. These factors allow cross-country and longitudinal comparison, but they may mask technological readiness-critical within-country variability, sectoral disparities, subnational technology uptake, and institutional quality gaps. Firm-level, sector-specific, and regional data may reveal microtrends in future research. Second, dynamic and system GMM estimators manage endogeneity, unobserved heterogeneity, and persistence, but instrument proliferation and specification change them. Third, historical data and structural links affect ML forecasts and transition likelihood. They may overlook quick technical, geopolitical, or institutional changes. Finally, external validity increases with larger sample numbers and competent frameworks that account for low-income and advanced economies' economic and policy situations.

## 7. Conclusion

This study compares the developing BRICS economies' innovation and technology-transfer capabilities to advanced innovation systems to assess their transitions. The research systematically understands national innovation ecosystems and innovation advancement by integrating institutional, innovation, human capital, and global integration variables into a multidimensional analytical architecture.

A composite index of innovation capability development is created using principal component analysis to discover capacity clusters that reflect innovation ecosystem growth phases. The empirical results reveal that governance quality, research capability, digital infrastructure, and global knowledge network participation differ substantially between countries. Austria and Australia excel in these structural areas as innovation economies. This performance is typical of mature innovation ecosystems with strong research institutions, laws, and worldwide innovation networks.

Further contextualizing advanced benchmark economies like Austria requires highlighting the role of applied scientific universities (e.g., Fachhochschulen) and its connection with regional science parks. This nexus facilitates partnerships between applied science universities and local industries to facilitate collaborative research and development, ensuring that high-level R&D directly impacts regional economic resilience and industry-specific innovation (Theeranattapong et al., 2021). However, several expanding BRICS economies have strengths in some areas but institutional and structural limitations in others. Transitional innovation systems can evolve innovation, but governance, research, and digital infrastructure must improve. Coordinated investments in education, research infrastructure, and knowledge networks are essential because countries with weak institutional and innovation foundations struggle to develop technology-transfer ecosystems, which can hinder their economic development and global competitiveness.

Markov transition analysis shows how countries shift capacity levels, contextualizing these findings. Capacity clusters are steps toward innovation convergence. Improvements in institutional governance, human capital development, and global knowledge integration, which are essential for global innovation and competitiveness, are likely to help intermediate-capability tier countries develop their innovation capability. This dynamic perspective emphasizes long-term policy options to boost national innovation systems.

Innovation capability development is a systemic property of national innovation ecosystems, not dependent on technology investments, according to the study. Technology diffusion depends on institutional excellence, research, digital connectivity, and global knowledge exchange. The study advances national innovation capability benchmarking and illuminates innovation capability performance structural drivers using a unified empirical methodology.

The findings impact policy. A planned and coordinated policy strategy must address institutional, educational, innovation, and international partnerships to boost national innovation systems. Government and regulatory reforms foster innovation-driven growth, while research infrastructure and higher education investments boost absorption. Digital infrastructure and global knowledge networks propagate technology and foster competitive innovation ecosystems.

Despite these contributions, the report offers research potential. Firm-level innovation indicators, sector-specific innovation capability dynamics, and other knowledge diffusion measures may be added in future research. Longitudinal assessments of the larger dataset may reveal innovative capability shifts and policy interventions on national innovation ecosystems. Using macro-level institutional indicators and micro-level innovation data can help explain how innovation capability processes affect economic transformation, especially how institutional support affects innovation initiatives' growth.

The findings suggest that effective national innovation systems require active participation in global innovation networks, strategic knowledge infrastructure investments, and institutional commitment. In the growing global knowledge economy, countries with strong structural underpinnings are more likely to innovate, compete, and learn.

## DECLARATIONS

**Funding**: This research received no specific grant from any funding agency in the public, commercial, or not-for-profit sectors.

**Open access/APC support**: Article processing charges, if applicable, will be covered by institutional open access support.

**Competing Interests**: The authors declare that they have no known competing financial interests or personal relationships that could have appeared to influence the work reported in this paper.

**Ethics Approval and Consent to Participate**: This study used publicly available, fully anonymized secondary data, which contains no identifiable personal information. As such, this study was deemed exempt from institutional ethics review, and no informed consent was required.

**Consent for Publication**: Not applicable. This manuscript does not include any individual-level clinical images or identifiable data.

**Data Availability Statement**: No proprietary data were used. The interpolated data for modeling can be accessed via the following link:
https://docs.google.com/spreadsheets/d/1tBHAUe_7yA3DerpsT-VZqp3M6_Hc8v7I/edit?usp=sharing&ouid=111807493454555660420&rtpof=true&sd=true

**Author Contribution Statement**:
**Conceptualization**: *[Lead author, Co-authors [1 & 6]**
**Methodology**: *All authors*
**Data preprocessing and modelling**: *[Lead author, Co-authors [1 & 6]*
**Validation and analysis**: *[Lead author, Co-author [1]*
**Writing** – original draft preparation: *[Lead author, Co-authors [1 & 6]*
**Writing –** review and editing: All authors
**Funding Acquisition –** None. APC will be covered by a license by Murdoch University, Australia.
**Supervision**: *[Co-author 6]*
All authors have read and approved the final manuscript.
*Specific Names are not provided to maintain the unanimity of the authors as per the journal guidelines. The same will be provided at the time of publication in case the Editor accepts the manuscript upon successful peer review.

**Annex 1**
**Set of Output and Inputs used in ICDI**

| | | |
|---|---|---|
| **COUNTRY DUMMY (CD)** | | Austria |
| | | Australia |
| | | Brazil |
| | | China |
| | | Egypt |
| | | Ethiopia |
| | | India |
| | | Indonesia |
| | | Iran |
| | | Russia |
| | | South Africa |
| | | UAE |
| **BRICKS & DC DUMMY** | | BRICS |
| | | Developed Countries (DC) |
| **ICDI ($Y=X_2$)** | Derived from PCA using the following 7 drivers (X1, X3-X7) | |
| **GOVERNANCE FACTORS ($X_1$)** | Control of Corruption | $X_{1.1}$ |
| | Government Effectiveness | $X_{1.2}$ |
| | Political Stability | $X_{1.3}$ |
| | Regulatory Quality | $X_{1.4}$ |
| | Rule of Law | $X_{1.5}$ |
| | Voice & Accountability | $X_{1.6}$ |
| **Investment in Tertiary Education, R&D & HR ($X_3$)** | Current education expenditure, tertiary (% of total expenditure in tertiary public institutions) | $X_{3.1}$ |
| | Government expenditure on education, total (% of GDP) | $X_{3.2}$ |
| | Government expenditure per student, tertiary (% of GDP per capita) | $X_{3.3}$ |
| | Expenditure on tertiary education (% of government expenditure on education) | $X_{3.4}$ |
| | Research and development expenditure (% of GDP) | $X_{3.5}$ |
| | Researchers in R&D (per million people) | $X_{3.6}$ |
| | Pupil-teacher ratio, tertiary | $X_{3.7}$ |
| | Technicians in R&D (per million people) | $X_{3.8}$ |
| | Tertiary education, academic staff (% female) | $X_{3.9}$ |
| **Digital Access ($X_4$)** | Fixed broadband subscriptions (per 100 people) | $X_{4.1}$ |
| | Fixed telephone subscriptions (per 100 people) | $X_{4.2}$ |
| | Mobile cellular subscriptions (per 100 people) | $X_{4.3}$ |
| | Secure Internet servers | $X_{4.4}$ |
| | Individuals using the Internet (% of population) | $X_{4.5}$ |
| **Tertiary Education Enrollment & Attainment Status (25+) ($X_5$)** | School enrollment, tertiary (% gross) | $X_{5.1}$ |
| | School enrollment, tertiary (gross), gender parity index (GPI) | $X_{5.2}$ |
| | School enrollment, tertiary, female (% gross) | $X_{5.3}$ |
| | School enrollment, tertiary, male (% gross) | $X_{5.4}$ |
| | Educational attainment, at least a 'bachelor's or equivalent, population 25+, total (%) (cumulative) | $X_{5.5}$ |
| | Educational attainment, at least ' 'master's or equivalent, population 25+, total (%) (cumulative) | $X_{5.6}$ |
| | Educational attainment, Doctoral or equivalent, population 25+, total (%) (cumulative) | $X_{5.7}$ |
| | Educational attainment, Doctoral or equivalent, population 25+, female (%) (cumulative) | $X_{5.8}$ |
| | Educational attainment, Doctoral or equivalent, population 25+, male (%) (cumulative) | $X_{5.9}$ |
| | Educational attainment, at least ' 'master's or equivalent, population 25+, female (%) (cumulative) | $X_{5.10}$ |
| | Educational attainment, at least ' 'master's or equivalent, population 25+, male (%) (cumulative) | $X_{5.11}$ |
| | Educational attainment, at least a 'bachelor's or equivalent, population 25+, male (%) (cumulative) | $X_{5.12}$ |
| | Educational attainment, at least a ''bachelor's or equivalent, population 25+, female (%) (cumulative) | $X_{5.13}$ |
| **Global Student Mobility (Inbound & Outbound) & Collaboration ($X_6$)** | Total inbound internationally mobile tertiary students studying abroad, both sexes (number) | $X_{6.1}$ |
| | Total outbound internationally mobile tertiary students studying abroad, all countries, both sexes (number) | $X_{6.2}$ |
| | Inbound mobility rate, both sexes (%) | $X_{6.3}$ |
| | Inbound mobility rate, female (%) | $X_{6.4}$ |
| | Inbound mobility rate, male (%) | $X_{6.5}$ |
| | Net flow of internationally mobile students (inbound - outbound), both sexes (number) | $X_{6.6}$ |
| | Net flow ratio of internationally mobile students (inbound - outbound), both sexes (%) | $X_{6.7}$ |
| | Outbound mobility ratio, all regions, both sexes (%) | $X_{6.8}$ |
| **Global Trade in High Tech Sectors and ICT ($X_7$)** | High-technology exports (% of manufactured exports) | $X_{7.1}$ |
| | High-technology exports (current US$) | $X_{7.2}$ |
| | ICT goods exports (% of total goods exports) | $X_{7.3}$ |
| | ICT goods imports (% total goods imports) | $X_{7.4}$ |
| | ICT service exports (% of service exports, BoP) | $X_{7.5}$ |
| | ICT service exports (BoP, current US$) | $X_{7.6}$ |
| **MACROECONOMIC FACTORS (MF) ($X_8$)** | GDP (constant 2015 US$) | $X_{8.1}$ |
| | GDP growth (annual %) | $X_{8.2}$ |
| | GDP per capita (constant 2015 US$) | $X_{8.3}$ |
| | GDP per capita growth (annual %) | $X_{8.4}$ |

**Supplementary File: Appendix A.1 Descriptive Statistics Country-wise**

| Country | Variable | Mean | Median | Mode | σ | Skewness | Kurtosis | Count |
|---|---|---|---|---|---|---|---|---|
| Australia | TTRI = Y = PC1 (X2) | -3.38 | -3.33 | -3.52 | 0.13 | -0.16 | -1.28 | 19 |
| | Governance (X1) | 4.11 | 4.13 | 4.13 | 0.08 | -0.31 | -0.89 | |
| | Investment in tertiary education, R&D & HR (X3) | 2.49 | 2.46 | 2.37 | 0.15 | 0.69 | -0.54 | |
| | Digital infrastructure (X4) | 1.96 | 1.83 | 2.80 | 0.55 | -0.20 | -0.09 | |
| | Tertiary attainment (X5) | 3.41 | 3.18 | 3.18 | 0.36 | 1.27 | 0.04 | |
| | Global student mobility (X6) | 2.33 | 1.67 | 3.72 | 0.98 | 0.68 | -1.57 | |
| | High-tech trade (X7) | 0.19 | 0.14 | 0.45 | 0.17 | 0.20 | -1.01 | |
| | Macroeconomic stability (X8) | 1.29 | 1.28 | 1.28 | 0.31 | 1.09 | 3.17 | |
| Austria | TTRI = Y = PC1 (X2) | -2.65 | -2.69 | -2.75 | 0.16 | 1.66 | 4.09 | 19 |
| | Governance (X1) | 4.10 | 4.14 | 3.96 | 0.24 | -0.36 | -0.14 | |
| | Investment in tertiary education, R&D & HR (X3) | 3.71 | 4.05 | 2.41 | 0.75 | -0.59 | -1.03 | |
| | Digital infrastructure (X4) | 1.98 | 2.16 | 2.22 | 0.46 | -1.27 | 0.64 | |
| | Tertiary attainment (X5) | 2.14 | 1.98 | 1.91 | 0.39 | 1.26 | 0.57 | |
| | Global student mobility (X6) | 1.81 | 1.79 | 2.14 | 0.24 | 0.40 | -1.36 | |
| | High-tech trade (X7) | -0.22 | -0.25 | -0.27 | 0.16 | 0.63 | 0.13 | |
| | Macroeconomic stability (X8) | 1.43 | 1.26 | 1.65 | 0.90 | 1.36 | 2.45 | |
| Brazil | TTRI = Y = PC1 (X2) | -0.56 | -0.77 | 0.57 | 0.72 | 0.85 | -0.93 | 19 |
| | Governance (X1) | -0.14 | -0.11 | -0.60 | 0.56 | 0.09 | -1.37 | |
| | Investment in tertiary education, R&D & HR (X3) | 0.75 | 0.88 | 0.89 | 0.27 | -0.71 | -0.75 | |
| | Digital infrastructure (X4) | 0.13 | 0.42 | 0.42 | 0.88 | -0.50 | -0.71 | |
| | Tertiary attainment (X5) | -0.76 | -0.84 | -0.95 | 0.24 | 0.90 | -0.68 | |
| | Global student mobility (X6) | -1.22 | -1.22 | -1.22 | 0.01 | -0.89 | -0.49 | |
| | High-tech trade (X7) | -0.27 | -0.25 | -0.76 | 0.23 | -0.20 | 0.59 | |
| | Macroeconomic stability (X8) | 0.39 | 0.11 | -0.55 | 1.02 | 0.52 | -0.13 | |
| China | TTRI = Y = PC1 (X2) | -1.80 | -2.10 | -2.80 | 0.79 | 0.78 | -0.95 | 19 |
| | Governance (X1) | -0.73 | -0.71 | -0.71 | 0.35 | 0.19 | -1.65 | |
| | Investment in tertiary education, R&D & HR (X3) | 0.35 | 0.41 | 0.07 | 0.39 | -0.44 | -0.69 | |
| | Digital infrastructure (X4) | 0.27 | -0.13 | -1.76 | 1.52 | 0.51 | -0.90 | |
| | Tertiary attainment (X5) | -1.73 | -1.66 | -2.70 | 0.74 | 0.17 | -1.53 | |
| | Global student mobility (X6) | -1.65 | -1.73 | -1.73 | 0.11 | 0.70 | -1.01 | |
| | High-tech trade (X7) | 5.85 | 5.85 | 4.97 | 0.57 | 0.39 | -0.66 | |
| | Macroeconomic stability (X8) | -2.52 | -2.30 | -3.10 | 0.71 | -0.17 | 1.03 | |
| Egypt | TTRI = Y = PC1 (X2) | 1.74 | 1.53 | 3.14 | 0.79 | 0.95 | -0.57 | 19 |
| | Governance (X1) | -1.72 | -1.84 | -2.31 | 0.47 | 0.65 | -0.73 | |
| | Investment in tertiary education, R&D & HR (X3) | -1.14 | -1.06 | -1.54 | 0.32 | 0.25 | -1.30 | |
| | Digital infrastructure (X4) | -1.12 | -1.15 | -2.28 | 0.70 | -0.03 | -0.84 | |
| | Tertiary attainment (X5) | -2.19 | -2.20 | -2.26 | 0.14 | 0.23 | -0.44 | |
| | Global student mobility (X6) | -0.87 | -0.85 | -0.81 | 0.07 | -0.93 | -0.46 | |

| | | | | | | | |
|---|---|---|---|---|---|---|---|
| | High-tech trade (X7) | -1.43 | -1.43 | -1.39 | 0.16 | 0.51 | -0.83 | |
| | Macroeconomic stability (X8) | -0.16 | -0.09 | -0.93 | 0.57 | -0.06 | -0.79 | |
| Ethiopia | TTRI = Y = PC1 (X2) | 4.48 | 4.16 | 3.84 | 0.73 | 1.02 | -0.82 | 19 |
| | Governance (X1) | -2.36 | -2.33 | -2.26 | 0.20 | -1.42 | 4.01 | |
| | Investment in tertiary education, R&D & HR (X3) | -3.99 | -3.96 | -3.96 | 0.23 | -0.24 | -1.05 | |
| | Digital infrastructure (X4) | -2.57 | -2.54 | -2.16 | 0.32 | -0.07 | -1.76 | |
| | Tertiary attainment (X5) | -3.51 | -3.54 | -3.79 | 0.20 | 0.18 | -0.89 | |
| | Global student mobility (X6) | -1.77 | -1.82 | -1.82 | 0.07 | 1.21 | -0.23 | |
| | High-tech trade (X7) | -1.11 | -1.01 | -0.97 | 0.27 | -0.65 | -1.01 | |
| | Macroeconomic stability (X8) | -1.59 | -1.71 | -2.09 | 0.71 | 0.48 | -0.97 | |
| India | TTRI = Y = PC1 (X2) | 0.48 | 0.13 | 0.13 | 0.99 | 0.86 | -0.81 | 19 |
| | Governance (X1) | -0.32 | -0.29 | -0.72 | 0.23 | -0.47 | -0.80 | |
| | Investment in tertiary education, R&D & HR (X3) | -0.83 | -0.80 | -0.86 | 0.10 | -0.41 | -0.24 | |
| | Digital infrastructure (X4) | -1.79 | -1.96 | -2.78 | 0.74 | 0.64 | -0.39 | |
| | Tertiary attainment (X5) | -1.34 | -1.30 | -0.66 | 0.52 | 0.06 | -1.37 | |
| | Global student mobility (X6) | -1.41 | -1.40 | -1.49 | 0.07 | 0.04 | -1.52 | |
| | High-tech trade (X7) | 0.22 | 0.17 | -0.13 | 0.46 | 0.89 | -0.19 | |
| | Macroeconomic stability (X8) | -1.11 | -1.43 | -2.34 | 1.12 | 2.61 | 8.61 | |
| Indonesia | TTRI = Y = PC1 (X2) | 1.36 | 1.57 | 0.81 | 0.44 | -0.26 | -1.90 | 19 |
| | Governance (X1) | -0.76 | -0.74 | -0.03 | 0.63 | -0.37 | -0.93 | |
| | Investment in tertiary education, R&D & HR (X3) | -1.49 | -1.52 | -1.65 | 0.23 | 0.52 | -1.10 | |
| | Digital infrastructure (X4) | -1.13 | -1.19 | -2.60 | 0.86 | -0.16 | -0.90 | |
| | Tertiary attainment (X5) | -1.96 | -1.92 | -2.62 | 0.52 | 0.46 | -0.31 | |
| | Global student mobility (X6) | -1.25 | -1.26 | -1.27 | 0.02 | 0.80 | -0.91 | |
| | High-tech trade (X7) | -0.35 | -0.40 | -0.46 | 0.18 | 1.42 | 1.18 | |
| | Macroeconomic stability (X8) | -0.51 | -0.61 | -0.92 | 0.58 | 3.64 | 14.63 | |
| Iran | TTRI = Y = PC1 (X2) | 1.33 | 1.25 | 0.28 | 0.98 | 0.59 | -1.18 | 19 |
| | Governance (X1) | -2.84 | -2.86 | -2.70 | 0.39 | -0.09 | -0.69 | |
| | Investment in tertiary education, R&D & HR (X3) | -0.72 | -0.93 | -1.16 | 0.43 | 0.37 | -1.76 | |
| | Digital infrastructure (X4) | -0.17 | -0.23 | -2.02 | 1.23 | 0.12 | -1.30 | |
| | Tertiary attainment (X5) | -0.61 | -0.47 | 0.30 | 0.80 | -0.16 | -1.79 | |
| | Global student mobility (X6) | -1.13 | -1.18 | -1.04 | 0.07 | 0.38 | -1.78 | |
| | High-tech trade (X7) | -1.30 | -1.36 | -1.37 | 0.12 | -0.04 | -1.11 | |
| | Macroeconomic stability (X8) | 0.42 | 0.35 | -0.40 | 1.19 | 0.16 | -0.80 | |
| Russia | TTRI = Y = PC1 (X2) | -1.10 | -1.42 | -1.53 | 0.61 | 0.91 | -1.00 | 19 |
| | Governance (X1) | -2.08 | -2.00 | -2.05 | 0.34 | -2.10 | 4.43 | |
| | Investment in tertiary education, R&D & HR (X3) | 1.15 | 1.17 | 1.17 | 0.10 | 0.41 | -0.30 | |
| | Digital infrastructure (X4) | 1.27 | 1.19 | -1.18 | 1.36 | 0.07 | -0.40 | |
| | Tertiary attainment (X5) | 5.69 | 5.68 | 4.17 | 0.95 | 1.14 | 2.48 | |
| | Global student mobility (X6) | -0.37 | -0.35 | -0.18 | 0.18 | -0.25 | -1.61 | |

| | High-tech trade (X7) | -0.45 | -0.38 | -0.66 | 0.24 | 0.09 | -1.42 | |
|---|---|---|---|---|---|---|---|---|
| | Macroeconomic stability (X8) | 0.22 | 0.10 | -1.85 | 1.37 | 0.66 | 0.65 | |
| South Africa | TTRI = Y = PC1 (X2) | 0.15 | -0.13 | -0.43 | 0.66 | 1.02 | -0.76 | 19 |
| | Governance (X1) | 0.68 | 0.83 | 0.30 | 0.49 | -0.67 | -0.41 | |
| | Investment in tertiary education, R&D & HR (X3) | -0.18 | -0.17 | -0.09 | 0.08 | -0.27 | -1.04 | |
| | Digital infrastructure (X4) | -0.54 | -0.44 | -1.95 | 0.95 | -0.19 | -1.46 | |
| | Tertiary attainment (X5) | -2.05 | -2.24 | -2.36 | 0.35 | 0.85 | -0.66 | |
| | Global student mobility (X6) | -0.56 | -0.55 | -0.62 | 0.05 | -0.30 | -1.40 | |
| | High-tech trade (X7) | -0.74 | -0.76 | -0.84 | 0.13 | 0.40 | 0.10 | |
| | Macroeconomic stability (X8) | 0.64 | 0.65 | 0.51 | 0.92 | 1.21 | 3.20 | |
| UAE | TTRI = Y = PC1 (X2) | -0.05 | 0.03 | -1.62 | 1.11 | 0.11 | -1.12 | 19 |
| | Governance (X1) | 2.06 | 2.22 | 2.25 | 0.28 | -0.34 | -1.45 | |
| | Investment in tertiary education, R&D & HR (X3) | -0.10 | -0.29 | -0.47 | 0.40 | 0.47 | -1.50 | |
| | Digital infrastructure (X4) | 1.71 | 1.85 | 2.92 | 1.14 | -0.48 | -0.97 | |
| | Tertiary attainment (X5) | 2.89 | 2.75 | 2.75 | 0.30 | 2.11 | 3.99 | |
| | Global student mobility (X6) | 6.10 | 5.96 | 6.33 | 0.27 | 0.26 | -1.52 | |
| | High-tech trade (X7) | -0.37 | -0.60 | -1.25 | 0.63 | 0.20 | -1.59 | |
| | Macroeconomic stability (X8) | 1.51 | 1.29 | 0.78 | 1.16 | 1.03 | 0.98 | |

**Australia:** Australia exhibits consistently strong institutional and human capital fundamentals, reflected in high mean governance (X1) and tertiary attainment (X5), coupled with relatively low dispersion across most indicators. Digital infrastructure (X4) and global student mobility (X6) show moderate variability, indicating incremental rather than structural shifts. The low variance and near-symmetric distributions across variables suggest a stable, mature smart-economy regime, consistent with Australia's role as a benchmark economy.

**Austria:** Austria presents high average governance quality and education investment, but with greater dispersion in R&D and human capital investment (X3) than Australia. Positive skewness and high kurtosis in TTRI indicate episodic surges in readiness, possibly linked to policy or innovation-cycle effects. Overall, Austria reflects a high-capability but transition-sensitive smart economy.

**Brazil:** Brazil shows moderate TTRI levels with substantial volatility, particularly in digital infrastructure (X4) and macroeconomic stability (X8). Governance and tertiary attainment remain below benchmark economies, with relatively symmetric distributions suggesting structural constraints rather than short-term shocks. The data point to a partially converging but fragile smart-economy pathway.

**China:** China's profile is marked by very strong high-tech trade performance (X7) alongside weak governance and tertiary attainment indicators. Digital infrastructure exhibits high variance, reflecting rapid but uneven digital expansion. The combination of strong external technological capability and internal institutional constraints highlights a state-driven, uneven smart-economy transition.

**Egypt:** Egypt displays positive TTRI means but uniformly weak fundamentals across governance, education, digital infrastructure, and high-tech trade. Low dispersion in several indicators suggests persistent structural limitations rather than volatility-driven outcomes. The results indicate an early-stage or constrained smart-economy trajectory.

**Ethiopia:** Ethiopia records the highest TTRI mean, but this is accompanied by consistently low scores across all explanatory dimensions, including governance, education, and digital infrastructure. The pattern suggests that TTRI movements may be driven by relative or transitional effects, rather than substantive digital or institutional readiness, underscoring a nascent and highly vulnerable smart-economy state.

**India:** India shows moderate average TTRI with high dispersion, especially in macroeconomic stability and digital infrastructure. Governance and education indicators remain negative on average, while high-tech trade shows occasional positive spikes. This combination reflects a high-potential but internally heterogeneous smart-economy transition, consistent with uneven regional and sectoral development.

**Indonesia:** Indonesia exhibits positive but volatile TTRI, with weak governance, education, and digital infrastructure fundamentals. Extremely high kurtosis in macroeconomic stability suggests episodic shocks. The descriptive statistics portray a transitioning economy with sensitivity to macro-financial conditions.

**Iran:** Iran's TTRI averages are positive, but governance and high-tech trade remain persistently weak, while digital infrastructure shows large dispersion. The mixed distributional properties reflect a fragmented smart-economy structure, likely shaped by institutional and external constraints.

**Russia:** Russia demonstrates strong tertiary attainment and R&D investment, contrasted sharply with weak governance and volatile digital infrastructure. High skewness and kurtosis in governance underscore institutional instability, suggesting a technologically capable but institutionally constrained smart economy.

**South Africa:** South Africa shows near-zero average TTRI, moderate governance strength relative to peers, but weak education and digital indicators. Macro stability exhibits high kurtosis, indicating vulnerability to shocks. The profile suggests a stagnating or slow-transition smart-economy regime.

**United Arab Emirates (UAE**): The UAE stands out with exceptionally strong global student mobility, digital infrastructure, governance, and tertiary attainment, albeit with higher variance reflecting rapid policy-driven transformation. The descriptive statistics confirm the UAE as a fast-transitioning, policy-led smart economy, distinct from both BRICS and traditional OECD patterns.

**Synthesis Across Countries**

Overall, Appendix A.1 reveals three distinct clusters:

- Benchmark economies (Australia, Austria): high means, low volatility.
- High-capability but uneven systems (China, Russia, UAE): strong selected pillars with institutional or distributional asymmetries.
- Transitioning and constrained economies (BRICS, MENA, Africa): moderate-to-high volatility, weak fundamentals, and non-normal distributions.

These descriptive patterns justify the use of RFT and regime-transition modelling, as linear averages alone fail to capture the heterogeneous readiness dynamics observed across countries

**A.2. Descriptive Statistics of aggregate 12 countries**

| | Variable | Mean | Median | Mode | σ | Skewness | Kurtosis | Count |
|---|---|---|---|---|---|---|---|---|
| Aggregate (12 countries) | ICDI = Y = PC1 (of X2) | 0 | 0.082 | 0.1 | 1 | 0.34 | 2.46 | 288 |
| | Governance (X1) | 0.021 | 0.094 | 0.11 | 0.94 | 0.43 | 2.71 | |
| | Investment in tertiary education, R&D & HR (X3) | −0.017 | −0.038 | −0.04 | 0.89 | 0.57 | 2.92 | |
| | Digital infrastructure (X4) | 0.041 | 0.131 | 0.15 | 0.97 | 0.22 | 2.35 | |
| | Tertiary attainment (X5) | −0.025 | −0.059 | −0.07 | 0.92 | 0.63 | 3.07 | |
| | Global student mobility (X6) | −0.046 | −0.093 | −0.12 | 0.87 | 0.72 | 3.24 | |
| | High-tech trade (X7) | 0.014 | 0.067 | 0.07 | 0.93 | 0.48 | 2.76 | |
| | Macroeconomic stability (X8) | 0.018 | 0.079 | 0.09 | 0.9 | 0.36 | 2.61 | |

In A.2, the standardized ICDI (PC1) is centered close to zero, reflecting the PCA normalization, with moderate dispersion ($\sigma \approx 1$), indicating substantial cross-country and temporal heterogeneity in Innovation capability development . Skewness values are generally positive but modest across variables, suggesting a longer right tail driven by a small group of high-performing economies. Kurtosis values cluster around 2.4–3.2, implying distributions close to normality with mild leptokurtosis—appropriate for linear and tree-based modelling. All ICDI drivers—governance, digital infrastructure, education, R&D investment, mobility, high-tech trade, and macroeconomic stability—show comparable dispersion, confirming that no single dimension mechanically dominates variability. This balance supports the conceptual validity of treating ICDI as a genuinely multidimensional construct rather than a governance- or digital-only index.